\documentclass[epj]{svjour}
\usepackage{graphics}
\usepackage{color}
\usepackage{graphicx}

\begin{document}

\title{Matter-wave scattering on a BEC in a double-well potential}
\author{Stefan Hunn$^1$, Moritz Hiller$^1$, and Andreas Buchleitner$^1$ \vspace{0.15cm}\\ \vspace{0.15cm}Doron Cohen$^2$ \\ Tsampikos Kottos$^3$}
\institute{Physikalisches Institut, Albert-Ludwigs-Universit{\"a}t Freiburg, Hermann-Herder-Str. 3, 79104 Freiburg, Germany
\and Department of Physics, Ben-Gurion University, Beer-Sheva 84105, Israel \and Department of Physics, Wesleyan
University, Middletown, Connecticut 06459, USA}
\date{Received: date / Revised version: date}
\abstract{
We study the inelastic scattering of a probe particle on a Bose-Einstein condensate confined in a double-well potential.
We identify prominent signatures of the underlying mean-field phase space in the scattering signal and derive an analytical
expression for the inelastic scattering cross section.
 % end of PACS codes
} %end of abstract
\authorrunning{Hunn, Hiller, Buchleitner, Cohen, and Kottos}
\maketitle
\section{Introduction}
Bose-Einstein condensed particles (BEC) trapped in optical potentials have sparked an ever growing research activity 
which has led to many beautiful results such as the direct observation of the Mott-insulator to superfluid phase transition \cite{Greiner_etal02}
and of Anderson localization of matter waves \cite{Billy_etal08,RDFFMMI08}.
More recently, hybrid quantum systems moved into focus, ranging from BECs coupled to micromechanical oscillators \cite{HCHKKRT10},
over integrated single-atom detectors based on field ionization from graphene nanotubes \cite{GJSVHKGF09}, to fluorescence imaging on atom chips \cite{HWRHS09}.
Interestingly, even the smallest (non-trivial) lattice, the double-well potential, bears rich physics like the appearance of macroscopic, quantum self-trapped states \cite{MCWW97,Self_trapping_ober} and resulting complex decay scenarios \cite{HKO06,GKN08,WFW08,RK09}, nonlinear Landau-Zener crossings \cite{WN00,MCHKC09,WGK06},
and also provides insight into seemingly unrelated effects, like e.g., its interpretation as a bosonic Josephson junction \cite{J86}.
Recent experiments even took advantage of the inter-atomic interaction present in such systems to generate squeezed, i.e. entangled, states \cite{EGWGO08}
that allow to approach the Heisenberg limit in atom interferometry \cite{Riedel:2010zg,Gross:2010ye}

The wealth of observed physical phenomena stems from the many-body nature of the bosonic system.
While the microscopic description of the double well is given by the Bose-Hubbard (BH) Hamiltonian,
a basic understanding can be gained by the discrete Gross-Pitaevskii equation (GPE), that constitutes the mean-field limit of the BH
Hamiltonian, i.e., it describes the double-well physics for large particle numbers.
In the GPE, the inter-atomic interactions lead to nonlinear equations of motion and, albeit being integrable \cite{SLE85},
the mean-field dynamics of a BEC in a double-well potential can become complex: 
In general, there are two distinct phase-space regions, corresponding to the experimentally observed \cite{Self_trapping_ober},
Rabi-like oscillations and self-trapping of the condensate, respectively.

In our present contribution, we investigate inelastic mat\-ter-wave scattering of a probe particle on a BEC confined in a double-well potential,
implementable, e.g., as a hybrid quantum system on an atom chip. 
We employ the quantum mechanical treatment of the scattering process that we introduced in Ref.~\cite{HHBCK10} to study scattering on a chaotic three-site trap.
The main question we will address, is whe\-ther and how the underlying phase-space structure is manifest in the quantum scattering signal.
While in Ref.~\cite{HHBCK10} the chaoticity of the target required a statistical analysis, we anticipate that due to its
integrability, the double-well dynamics leaves {\em immediate} fingerprints in the scattering quantities.

This paper is structured as follows: In the next section we introduce the target system, a BEC in a double
well, and discuss its mean-field dynamics and corresponding quantum properties. Secs.~\ref{sec.setup} and
\ref{sec.S-matrix} are devoted to the scattering setup and its formal description via the scattering matrix, respectively.
Our results are presented in Sec.~\ref{sec.results}.

%%%%%%%%%%%%%%%%%%%%%%%%%%%%%%%%%%%%%%%%%%%%%%%%%%%%%%%%%%%%%%%%
%%%%%%%%%%%%%%%%%%%%%%%%%%%%%%%%%%%%%%%%%%%%%%%%%%%%%%%%%%%%%%%%

\section{Model}
\subsection{Scattering target}
\label{sec.target}
The scattering target is defined by $N$ ultra-cold bosons in a double-well potential,
described by the Bose-Hubbard (BH) Hamiltonian \cite{MCWW97,JBCGZ98}:
\begin{equation}
H_{BH} = \frac{U}{2}\sum_{j=1}^2 \hat{n}_j(\hat{n}_j - 1) -  k  ( \hat{b}_1^{\dagger} \hat{b}_2 +  \hat{b}_2^{\dagger} \hat{b}_1 )  \, . 
\label{eq.BHH}
\end{equation}
Here, $\hat{b}_j^{(\dagger)}$ are the bosonic annihilation (creation) operators, and $\hat{n}_j=\hat{b}_j^\dagger\hat{b}_j$ is the number counting
operator at site $j$.%
\footnote{\label{foot.bias}Numerically, we add a small bias of the order of $10^{-2}\hat n_1$, in order to break the symmetry of Eq.(\ref{eq.BHH}) and thereby
avoid unstable macroscopic superposition states \cite{Carr}.} 
$U$ and $k$ parameterize the on-site interaction and the tunneling strength, respectively. Experimentally, both parameters
can be independently controlled via the height of the potential barrier and by additional magnetic fields \cite{MO06} that induce
Feshbach resonances. 
Apart from the  total energy $E$, also the total particle number $N$ is a constant of motion.
The set of Fock basis states is thus given by $\{|n_1\rangle\}$, where $n_1\in \{0,1,...,N\}$ denotes
the boson number in well one, resulting in a Hilbert space of dimension $N+1$.

It is advantageous to introduce the angular momentum operators \cite{MCWW97}:
\begin{eqnarray}
\hat{J}_x &=& \   (\hat{b}_2\hat{b}_1^{\dagger} + \hat{b}_1\hat{b}_2^{\dagger})/2 \nonumber \, , \\
\hat{J}_y &=&  i (\hat{b}_2\hat{b}_1^{\dagger} - \hat{b}_1\hat{b}_2^{\dagger})/2 \, , \\
\hat{J}_z &=&  \  (\hat{b}_2\hat{b}_2^{\dagger} - \hat{b}_1\hat{b}_1^{\dagger})/2 \nonumber \, ,
\label{eq.Bloch}
\end{eqnarray}
which obey the commutation relation of an {\em su}(2) Lie-algebra.
In that representation, Hamiltonian (\ref{eq.BHH}) can be rewritten as $H_{BH} =  U\hat{J}_z^2 - 2k\hat{J_x}$, up to
a constant term. The conservation of the total boson number $N$ corresponds to $[\hat{J}^2, H_{BH}]=0$ and
the physical interpretation of the operators $\hat{J}_i$ is as follows: $\hat{J}_z$ measures the particle imbalance between the
wells, while $\hat{J}_y$ represents the condensate's momentum, and $\hat{J}_x$ bears direct information about the relative phase
of the condensate's fractions in the left and right well.

In the mean-field limit (corresponding to large particle numbers $N$), the dynamics of the condensate is described by the discrete
Gross-Pitaevskii equation \cite{PS03}: The quantum operators $b_{j}^{(\dagger)}$ are then replaced by time-dependent, complex
amplitudes $A_{j}^{(*)}$, leading to the Hamiltonian:
\begin{equation}
{\cal H}_{\rm GP}/N  = \frac{UN}{2} \sum_{j=1}^2 |A_j|^4  - k  \left[A_1^* A_{2} + A_2^{*}A_1 \right]  \, .
\label{eq.GP}
\end{equation}
The $A_{j}$ obey the canonical equations $i\partial A_j/\partial t=
\partial{\cal H}_{\rm GP}/\partial A_j^*$, that yield equations of motion for four real variables. 
Subtracting the two constants of motion, the mean-field dynamics  effectively acts on a two-dimensional phase space and can
be mapped on a spin $N/2$ Bloch sphere, via a 
transformation to classical angular momentum variables $J_i$ (obtained from (\ref{eq.Bloch}) after replacing 
the operators $\hat{b}_i$ by the mean-field amplitudes $A_i$).

\begin{figure}
\begin{center}
\includegraphics[width=0.45\columnwidth,keepaspectratio]{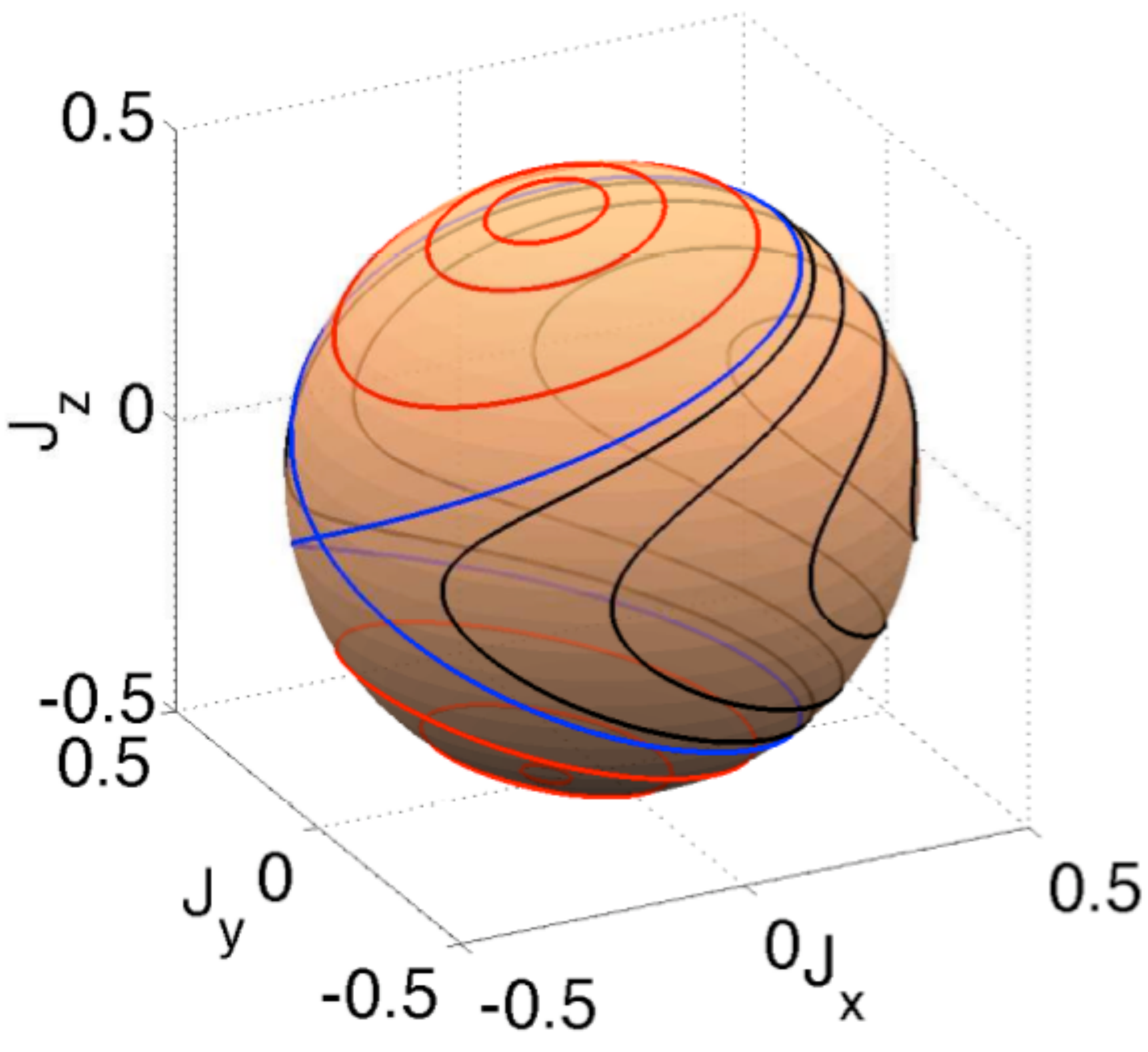}\hfill
\includegraphics[width=0.51\columnwidth,keepaspectratio]{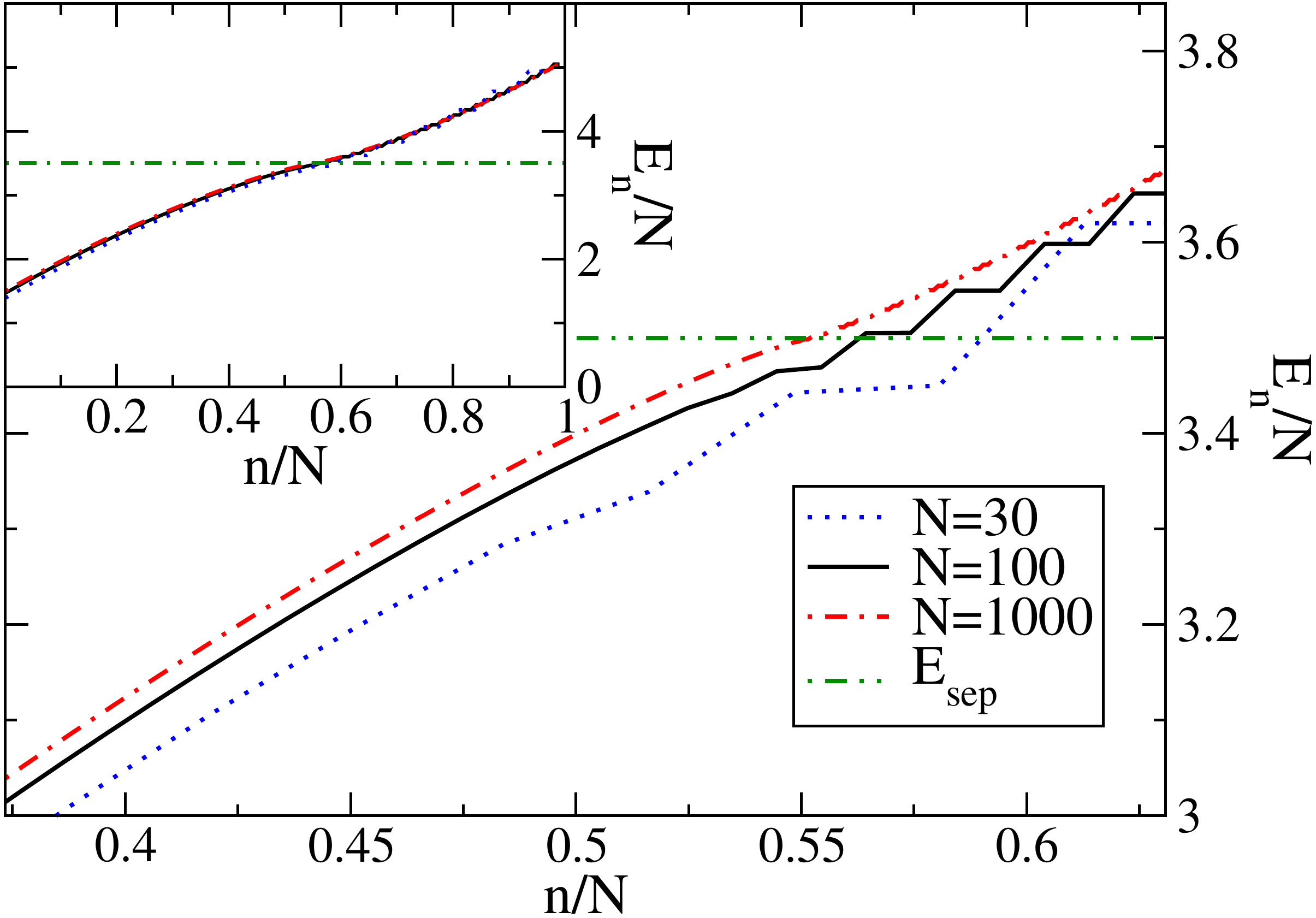}
\end{center}
\caption{(color online). 
{\it Left:} Mean-field phase space of a BEC in a double-well potential given by the discrete Gross-Pitaevskii equation (\ref{eq.GP}), for $u=5$.
Self-trapping modes (red) coexist with Rabi-like oscillations (black). The two regions are separated by the separatrix (blue).
There are two elliptic fixed points associated with self-trapped solutions.
{\it Right:} The rescaled energy spectra $E_n/N$ of the BH Hamiltonian for various particle numbers $N$, and constant control parameter $u=5$.
The separatrix energy $E_{\rm sep}$ is plotted as dash-double-dotted, horizontal, green line.
}
\label{fig.separatrix}
\end{figure}

The dynamics generated by (\ref{eq.GP}) is governed by the control parameter $u=UN/2k$ \cite{BES90}:
For very small inter-particle interaction $u < 1$, the BH system is in the Rabi regime, where the condensate oscillates between the 
two wells, i.e., the time-averaged population imbalance between them is zero (i.e., $\langle J_z\rangle_t = 0$). 
In the other extreme of very strong interactions $u>N^2$,
the system is in the self-trapped regime, where the condensate is persistently trapped in one of the wells, i.e., the average population imbalance
between them is non-zero $\langle J_z\rangle_t \neq 0$.

For intermediate values of the control parameter $u\approx 1$ (Josephson regime) the phase space is
divided by a separatrix located at the energy $E_{\rm sep}/N = k(u/2+1)$, and both
the above dynamical modes coexist \cite{SLE85,BES90,ELS85}.
In the left panel of Fig.\ref{fig.separatrix}, we show the phase-space structure typical for this intermediate regime:
For small initial particle imbalances (and phase difference $J_x\approx0.5$) Rabi-like oscillations (black) rule the phase-space.
Here, each trajectory corresponds to a situation where a fraction of the condensate oscillates between the two wells,
while, on average, the bosons are equidistributed, i.e., $\langle J_z\rangle_t=0$.
In this {\em Rabi region} of phase space and for a given energy, there exists exactly one trajectory.

In contrast, for sufficiently large initial population imbalance, one observes self-trapped trajectories (red).
They encircle the stable fixed points located in the Northern (Southern) hemisphere and correspond
to a persistent particle imbalance $J_z(t) > 0$ ($J_z(t) < 0$) for all times $t$.
For a given energy in the {\em self-trapped region} of phase space, a pair of two solutions with inverse particle imbalance exists.
The blue line denotes the separatrix.

This phase-space structure has a clear signature in the quantum energy spectrum of Hamiltonian (\ref{eq.BHH})
(cf. right panel of Fig.~\ref{fig.separatrix}): Energy eigenstates below the separatrix energy $E_{\rm sep}$ are non-degenerate, since 
their mean-field counterpart are uniquely defined trajectories in the Rabi region. 
The appearance of two self-trapped trajectories of the same energy is quantum mechanically reflected 
in the fact that energy eigenstates above the separatrix energy are pairwise
degenerate.%
\footnote{More precisely, the eigenstates of (\ref{eq.BHH}) with energy above $E_{\rm sep}$ 
are quasi-degenerate \cite{Carr,AFKO96} what is lifted by the small on-site potential (see Footnote~\ref{foot.bias}).
Nevertheless, the self-trapped states can be considered as pairwise degenerate, since their splitting is very small compared to all  
other scales in the spectrum of (\ref{eq.BHH}).}

In the present paper, we concentrate on the Josephson regime of intermediate interactions ($u=5$), and large numbers of bosons in
the double-well potential ($N\ge 30$).
In the following, eigenstates whose mean-field counterpart are trajectories that show Rabi-like (self-trapping) behavior
are called eigenstates in the Rabi (self-trapped) {\em region} of the spectrum.

We note that by construction (cf. Footnote~\ref{foot.bias}) none of the eigenstates of (\ref{eq.BHH}) corresponds to a macroscopic superposition state.
The creation of such states, their interaction with an electromagnetic field, and the resulting decoherence
properties have been studied in detail by \cite{HM06}.

%%%%%%%%%%%%%%%%%%%%%%%%%%%%%%%%%%%%%%%%%%%%%%%%%%%%%%%%%%%%%%%%

\subsection{Scattering setup}
\label{sec.setup}

\begin{figure}
\includegraphics[width=0.88\columnwidth,keepaspectratio]{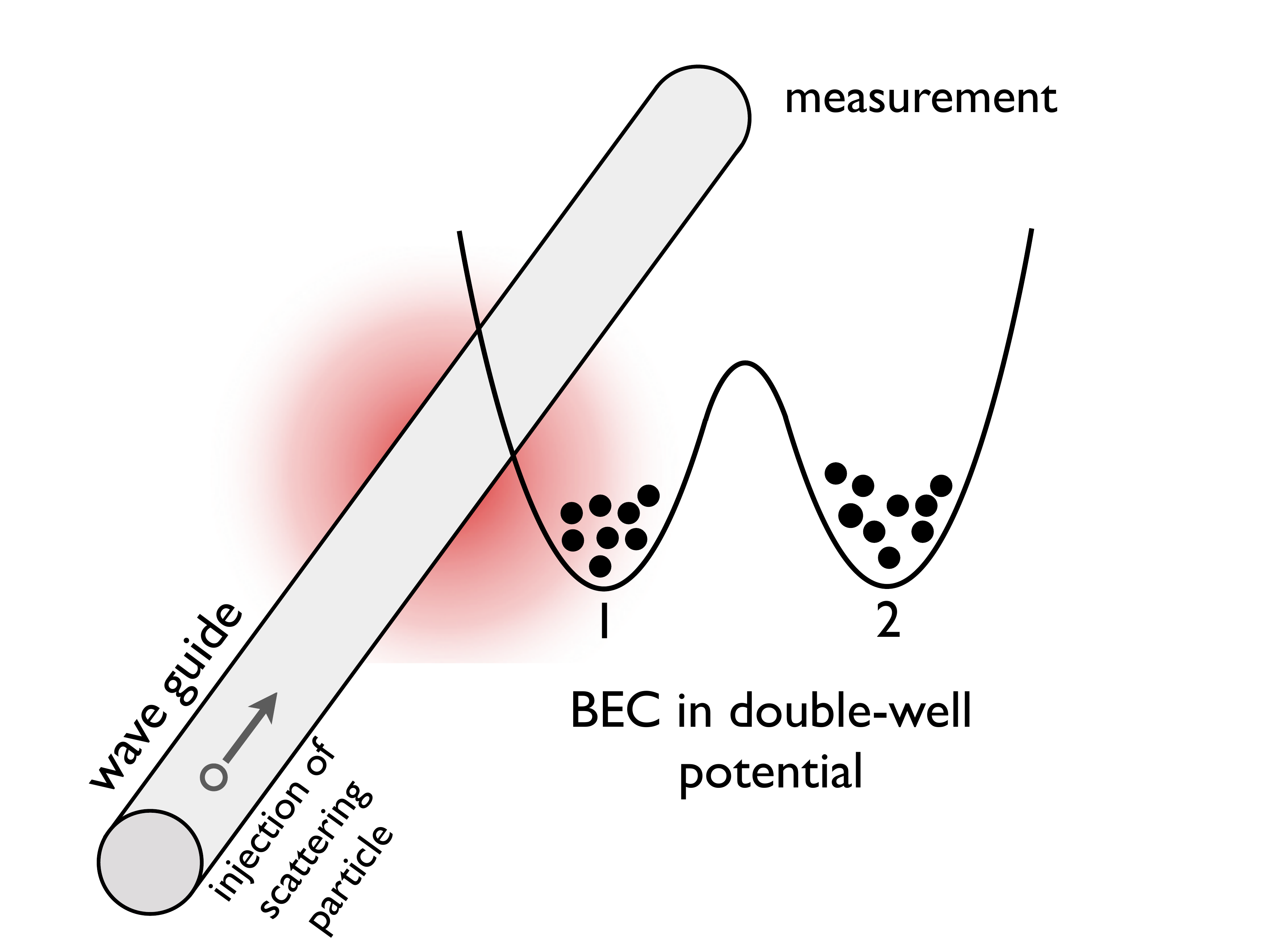}
\caption{(color online). Scattering setup: The probe particle is injected into a waveguide that passes a condensate confined by a
double-well potential. In the contact region between waveguide and site one of the double well, the probe particle and the
BEC can exchange energy. The scattering quantities of the particle measured on exit  from the waveguide carry detailed information on the
state of the condensate.
}
\label{fig.setup}
\end{figure}

We probe the double-well system by means of the scattering setup shown in Fig.~\ref{fig.setup} \cite{HHBCK10}: 
A probe particle with momentum $k$ moves in a waveguide that is placed in the proximity of say, site one of the double well. 
When the particle approaches the condensate, it interacts with the latter leading to an exchange of energy. On exit
from the waveguide, typical scattering quantities of the probe particle, like e.g., its inelastic scattering probability, are measured.
A possible experimental implementation of such scenario is provided by atom-chip setups, which allow for both,
trapping of the BEC and (magnetic) guiding of the probe particle.

The waveguide in our scattering scheme is modeled by two semi-infinite tight-binding (TB) leads with hopping term $J$ and 
lattice spacing $a=1$. These two leads are coupled with strength $J_0$ to the central lead site $j=0$, which is closest to the condensate.
Thus, $J_0 \le J$ determines the effective coupling strength between the leads and the projectile-target interaction region (i.e. for $J_0=0$ the latter
is completely isolated) and hence controls the width of the scattering resonances.
Consequently, the probe-particle Hamiltonian reads:
\begin{equation}
\label{Hwaveguide} 
H_{\rm TB} = \Big[ -J \sum_{j\ne -1,0} \hat{c}_j \hat{c}_{j+1}^{\dagger}
-  J_0 \sum_{j=-1,0} \hat{c}_j \hat{c}_{j+1}^{\dagger} \Big] + {\rm h.c.} \, ,
\end{equation}
with $\hat{c}_j^{(\dagger)}$ the annihilation (creation) operators of the probe particle 
at site $j$ of the TB lead. The particle's energy in the momentum eigenstate $|k_m\rangle$ is $\epsilon_m 
=-2 J \cos(k_m)$, with corresponding velocity $v_m = 2 J \sin(k_m)$ \cite{Datta95}.

The probe-target interaction $H_{\rm int}$ is assumed to be of similar type (i.e. short range) as the bosonic
inter-particle interaction in the condensate. Hence, it takes non-vanishing values only when the probe particle is located at the 
central lead site ($j=0$) which is closest to site one of the double well, and it is proportional to the number
of bosons at this site:
\begin{equation}
\label{eq.H_int}
H_{\rm int} =  \alpha \cdot \hat{c}^{\dagger}_0 \hat{c}_0 \otimes \hat{n}_1\, ,
\end{equation}
$\alpha\!>\!0$ is a parameter that controls the strength of the probe-target interaction.

%%%%%%%%%%%%%%%%%%%%%%%%%%%%%%%%%%%%%%%%%%%%%%%%%%%%%%%%%%%%%%%%

\subsection{Scattering matrix} 
\label{sec.S-matrix}
Given the total Hamiltonian 
\begin{equation}
H_{\rm tot} = H_{\rm TB} \otimes {\hat 1} + {\hat 1} \otimes H_{\rm BH} + H_{\rm int} \, ,
\label{eq.H_tot}
\end{equation}
we can now define the scattering matrix of our problem: 
The condensate is initially prepared in an energy eigenstate $|E_m\rangle$, while the probe particle is injected with an energy 
$\epsilon_m$. Hence, asymptotically far from the interaction region,
the total energy is ${\cal E}=E_m + \epsilon_m$.\footnote{In our calculations we rescale the BH spectrum (and the interaction operator) to lie
within the bandwidth of the lead, and thereby avoid evanescent modes.} 
The open channels (modes) in the leads are then determined by energy conservation and 
the fact that the condensate's final state is among the energy eigenstates $\{E_n\}$ of the BH Hamiltonian. 
Hence, the open modes are 
characterized by the kinetic energy $\epsilon_{n}={\cal E}-E_{n}$ of the outgoing probe particle.
In other words, for a given total energy ${\cal E}$, the basis of our scattering problem is completely characterized by the Bose-Hubbard
eigenstates, i.e. $\{  |\epsilon_n\rangle \otimes |E_n\rangle \} \equiv \{ |E_n\rangle\}$.
The transmission block of the $S$-matrix then reads \cite{HHBCK10}:
\begin{equation}
[{\hat S}_T]({\cal E}) =\sqrt{\hat v} \frac{i\,\gamma}{(1-\gamma)  [{\cal E}- {\hat H}_{BH}] - \alpha {\hat n}_1 + i\gamma {\hat v} } 
\sqrt{\hat v} \, ,
\label{eq.S_T}
\end{equation}
where $\gamma \equiv(J_0/J)^2$, and ${\hat v}$ is the velocity operator. 
In the eigenbasis of the BH Hamiltonian, both ${\hat H}_{BH}$ and ${\hat v}$ become diagonal matrices i.e. $[H_{BH}]_{nm}=E_n\delta_{nm}$
and $v_{nm}=v_n \delta_{nm}$. In contrast, $Q_{nm} = \langle E_n|{\hat n}_1|E_m\rangle$ is, in general ($k\neq0$), a non-diagonal
matrix, what yields a non-diagonal $S_T$-matrix. This, in turn, corresponds to inelastic scattering and in that sense $\alpha$ controls the degree
of inelasticity in the scattering process. 

As known from scattering theory, the imaginary part of the $S$-matrix denominator determines the width of the resonances.
Thus, in the present case, the coupling ratio $\gamma$ controls this width:
For example, in the limit $\gamma \rightarrow 0$, the resonances become $\delta$-like, while they grow with increasing $\gamma$.
For $\gamma=1$, Eq.~(\ref{eq.S_T}) coincides with the $S$-matrix for inelastic electronic scattering in a 1D geometry derived in \cite{BC08}.
Throughout the paper, we fix $\gamma=0.1$, corresponding to the intermediate regime of overlapping resonances.

%%%%%%%%%%%%%%%%%%%%%%%%%%%%%%%%%%%%%%%%%%%%%%%%%%%%%%%%%%%%%%%%
%%%%%%%%%%%%%%%%%%%%%%%%%%%%%%%%%%%%%%%%%%%%%%%%%%%%%%%%%%%%%%%%

\section{Results}
\label{sec.results}
\subsection{Interaction matrix $Q$}
\begin{figure}
\begin{center}
\includegraphics[width=0.46\columnwidth,keepaspectratio]{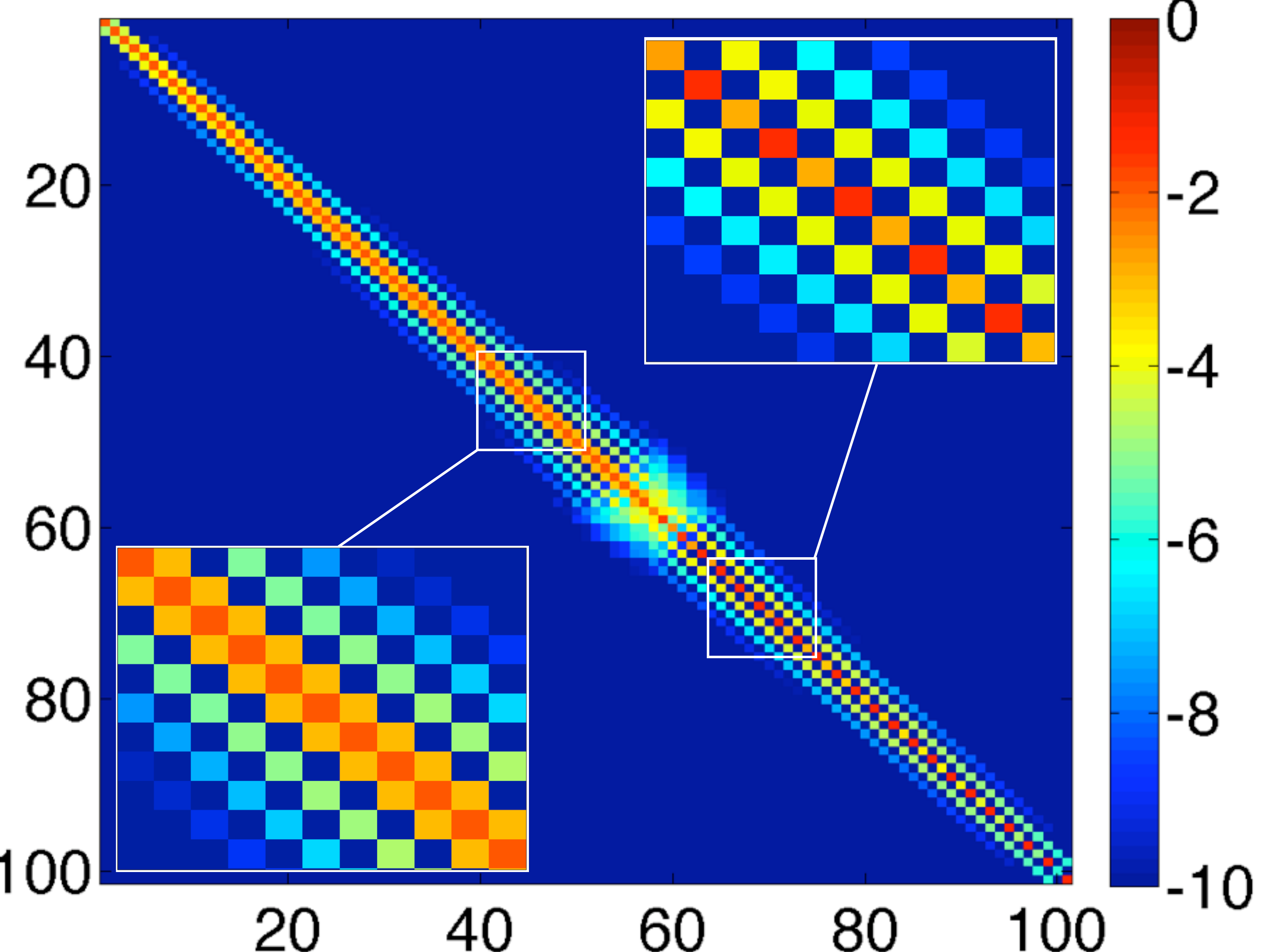}\hfill
\includegraphics[width=0.49\columnwidth,keepaspectratio]{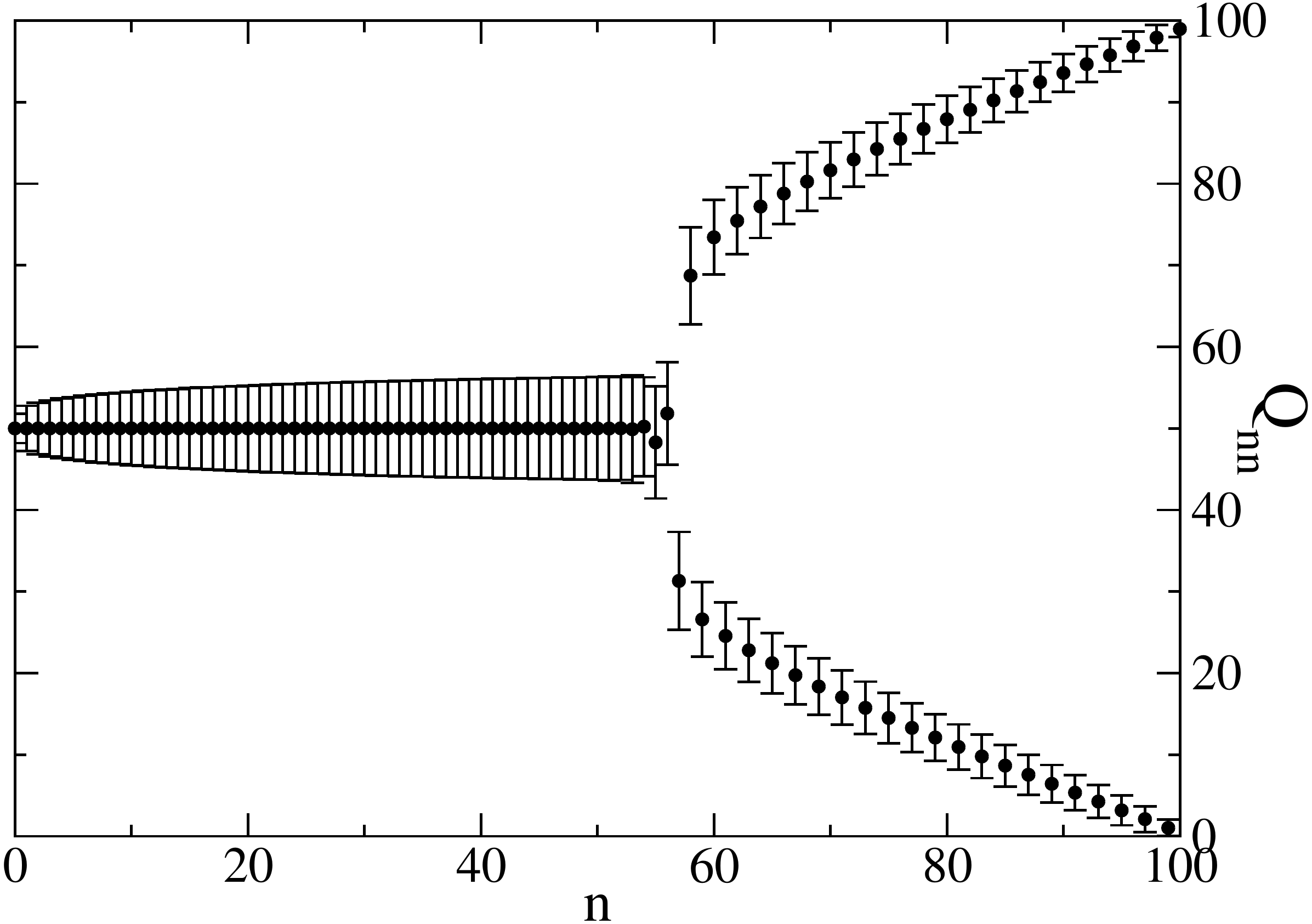}
\end{center}
\caption{(color online). 
{\it Left:} Logarithmically color-coded snapshot of the interaction matrix $Q$, for $N=100$ particles and $u=5$. The insets show
magnifications of the $Q$-matrix structure in the Rabi (left) and self-trapped (right) regime, respectively.
{\it Right:} The expectation value $\langle E_n|\hat{n}_1|E_n\rangle = Q_{nn}$ of the bosonic number operator $\hat{n}_1$
versus the level index $n$, for identical parameters. The error bars denote the corresponding standard
deviation $\sigma_n = \sqrt{\langle E_n|\hat{n}^2_1|E_n\rangle  - \langle E_n| \hat{n}_1|E_n\rangle^2}$.
}
\label{fig.Q-matrix}
\end{figure}

We start our analysis by a direct inspection of the probe-target interaction matrix $Q$ shown in the left panel of
Fig.~\ref{fig.Q-matrix}, for $N=100$ bosons.
The value of the control parameter is $u=5$, what corresponds to intermediate inter-atomic interaction strengths in the BEC (see Fig.~\ref{fig.separatrix})
and will be fixed throughout the paper.
One immediately recognizes that $Q$ is a banded matrix that is divided into two halves by a ``blurred" spot, located in the vicinity of level
$n = 56$. 
The behavior of the matrix elements in either region is quite different: Consider, for example, the diagonal elements
$Q_{n n} = \langle E_n | {\hat n}_1 | E_n \rangle$ that represent the expectation value of the boson number in well one,
with respect to the energy eigenstate $|E_n\rangle$ (see full dots in the right panel of Fig.~\ref{fig.Q-matrix}).
Below the energy level $n=56$, $Q_{nn}$ takes the constant value $N/2$, what corresponds to equidistribution of the bosons
among the two wells.
Above this energy level, the expectation value of $\hat{n}_1$ begins to oscillate, e.g., in state $n=57$ only $31$ particles occupy site one,
while in state $n=58$, the majority of $69$ bosons populates site one, and so on.

Besides the average value $Q_{nn}$, the corresponding standard deviation
$\sigma_n = \sqrt{\langle E_n|\hat{n}^2_1|E_n\rangle  - \langle E_n| \hat{n}_1|E_n\rangle^2}$
bears complementary information on the number fluctuations and is represented in Fig.~\ref{fig.Q-matrix} by the error bars.
One clearly sees that $\sigma_n$ is largest in the vicinity of level $=56$ and takes its smallest values for the maximally localized states. 

The observed behavior is a direct signature of the underlying mean-field dynamics of the BEC:
We recall from the discussion in Sec.~\ref{sec.target}, that for $N = 100$ bosons, the separatrix energy $E_{\rm sep}$ is located
around energy level $E_{56}$ (see black solid curve of Fig.~\ref{fig.separatrix}), i.e., around the same level at which the
expectation value of $\hat{n}_1$ starts to oscillate.
Hence, eigenstates $\{|E_n\rangle\}$ with expectation value $Q_{nn}=N/2$ correspond to those mean-field solutions
that show, on average, vanishing particle imbalance $\langle J_z\rangle_t=0$, and thus belong to the Rabi region of the spectrum.
The associated number fluctuations grow with increasing level index and take their maximum value close to the
separatrix, where the mean-field oscillations in the $J_z$-direction of phase space are largest (see left panel of Fig.~\ref{fig.separatrix}).
On the other hand, states with index $n>56$ are alternatingly localized on either well and thus belong to the self-trapped region.
Their fluctuations are reduced with increasing index $n$. In the mean-field phase space, increasing $n$ corresponds to
trajectories that are closer and closer to the two elliptic fixed points, where the oscillations in any direction are zero.

Note also the checker-board structure in the interaction matrix $Q$ that is present in the self-trapped region and which implies that
the interaction operator can solely induce transitions between self-trapped states located at the same well.
In contrast, in the Rabi region, the odd off-diagonals are strongly populated, while the even off-diagonals vanish.
This is due to the fact that the operator $\hat{n}_1$ does not commute with the parity operator ${\hat P}$, and can solely induce
transitions between states with different parity (see Appendix A).
Since energy eigenstates in the Rabi region of the spectrum have alternating parity, the even off-diagonal elements, which couple
eigenstates with {\em same} parity, are zero.%
\footnote{Although this argument strictly holds only for symmetric double wells, we have checked that a small on-site bias does not affect the results.}
In both regimes, the value of the off-diagonal matrix elements decreases exponentially with the distance from the diagonal. 

%%%%%%%%%%%%%%%%%%%%%%%%%%%%%%%%%%%%%%%%%%%%%%%%%%%%%%%%%%%%%%%%

\subsection{Participation number of $S_T$}
\label{sec.PN}
How is the structure of $Q$, that encodes the properties of the BEC, reflected in the scattering signal? 
In the following, we will focus on the {\em inelastic} part of the scattering signal that, in contrast to the
elastic part, accounts for all final configurations of the target, and thus bears much more detailed information.
Consequently, we analyze the off-diagonal elements of the $S_T$-matrix.

\begin{figure}
\centerline{\includegraphics[width=0.88\columnwidth,keepaspectratio]{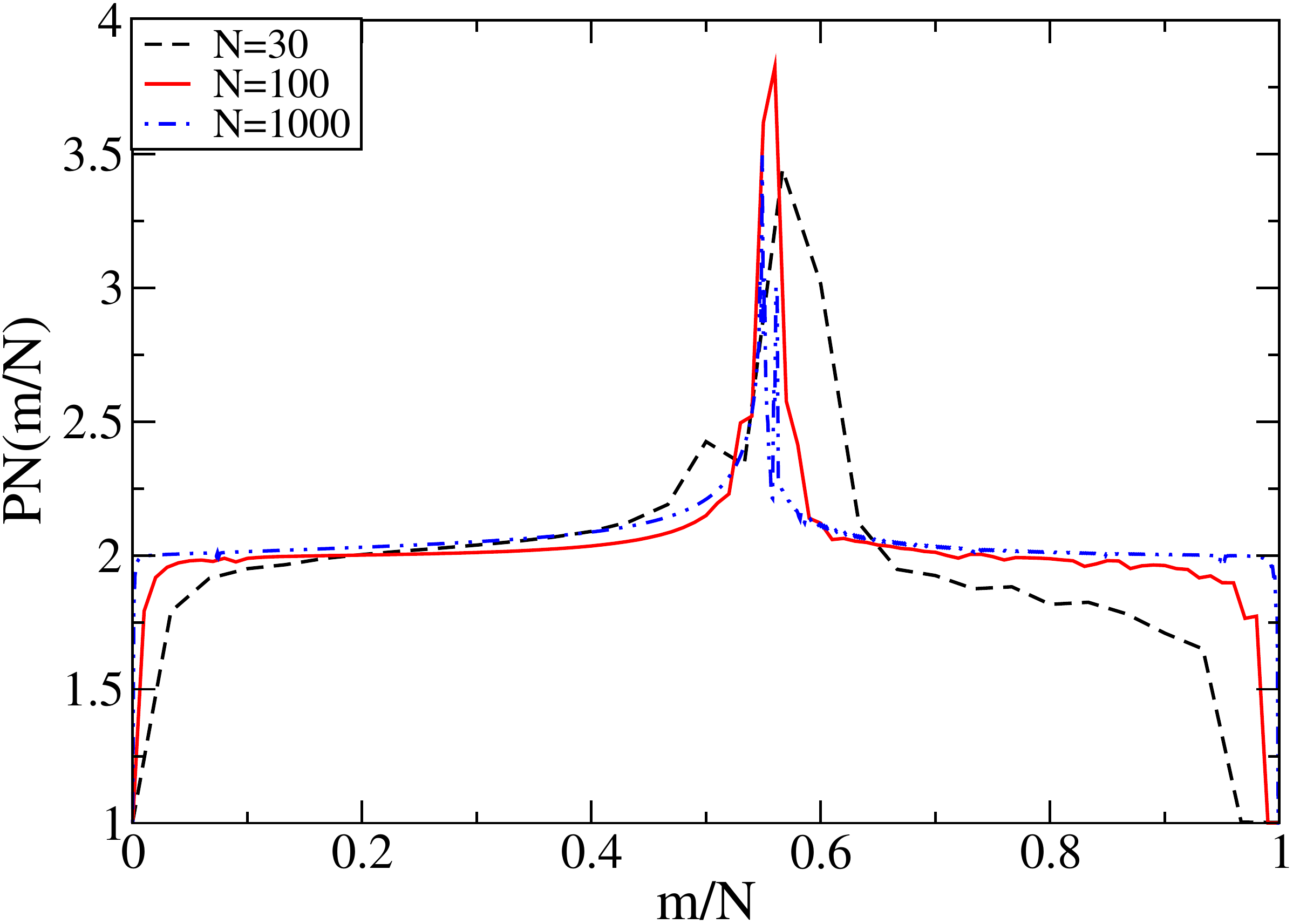}}
\caption{(color online). 
Participation number PN of the $S_{T}$-matrix versus the scaled level index $m/N$, for $u=5$, $\gamma=0.1$, $\alpha=1.0$, and
various boson numbers $N$. The PN is averaged over the entire accessible energy axis ${\cal E}$.
The peak around $m/N\approx0.56$ reflects the separatrix (compare right panel of Fig.~\ref{fig.separatrix}).
}
\label{fig.PN}
\end{figure}

As a first step, we determine the number of final target configurations $|E_n\rangle$ for the BEC initially
prepared in an eigenstate $|E_{m} \rangle$ ($m\neq n$).
To this end, we define the {\em participation number} PN of each column of the $S_T$ -matrix:
\begin{equation}
{\rm PN}(m) =  \left[ \frac{\sum_{n \neq m}  |[S_{\rm T}]_{n m}|^4 }{[\sum_{n\neq m} |[S_{\rm T}]_{n m}|^2]^2} \right]^{-1} \, .
\label{eq.PN}
\end{equation}
Since the diagonal term $[S_{\rm T}]_{mm}$ is excluded from the summation, this quantity is meaningful only if
the inelastic part of the scattering signal does not vanish, i.e. if $\alpha>0$.
Then, PN reflects the number of outgoing channels that participate in the inelastic process, and hence
takes values between one and $N$.

In Fig.~\ref{fig.PN}, we plot the PN for intermediate probe-target interaction $\alpha=1$ and various boson numbers
$N$, versus the scaled level index $m/N$.
All curves approximately assume the value two, apart from the strong peak in the PN located around $m/N = 0.56$.
Recalling the above discussion, we relate its position to the separatrix energy $E_{\rm sep}$.
Accordingly, for an initial preparation of the BEC in the self-trapped or Rabi region (i.e. $m/N \neq 0.56$), mainly two outgoing channels
take part in the inelastic scattering, while this number is enhanced for initial preparations near the separatrix energy $E_{\rm sep}$.

The observed behavior is a consequence of the shape of the interaction matrix $Q$ (see Fig.~\ref{fig.separatrix})
that can be understood from a semiclassical argument:
In the self-trapped and Rabi regions, the mean-field counterpart of the interaction operator, $n(t)=|A_1(t)|^2$, oscillates with one unique frequency \cite{CSHKVC10}, what, upon quantization, yields one transition frequency, i.e., two off-diagonals in the $Q$-matrix. 
In contrast, close to the separatrix, the two distinct dynamical behaviors approach. 
Therefore, in a small energy window around $E_{\rm sep}$, several frequencies appear in the dynamics of $n(t)$,
what leads to a larger number of off-diagonal elements in the interaction matrix $Q$.

%%%%%%%%%%%%%%%%%%%%%%%%%%%%%%%%%%%%%%%%%%%%%%%%%%%%%%%%%%%%%%%%

\subsection{Inelastic scattering cross section}
\label{sec.rho_in}
Besides the participation number PN of the $S_{\rm T}$-matrix, an experimentally
easily accessible quantity is the inelastic scattering cross section
\begin{equation}
\rho_{\rm in}^m({\cal E}) = 2 \sum_{n\neq m} |[S_T]_{nm}|^2 \, ,
\label{eq.rho_in}
\end{equation}
which denotes the probability to be scattered (forward and backward) from an incident channel $|\epsilon_m\rangle$ to any
other channel  $|\epsilon_n\rangle \neq |\epsilon_m\rangle$, or, equivalently, from a initially prepared condensate eigenstate $|E_m\rangle$
to a final eigenstate $|E_n\rangle$.

\begin{figure}
\centerline{\includegraphics[width=0.9\columnwidth,keepaspectratio]{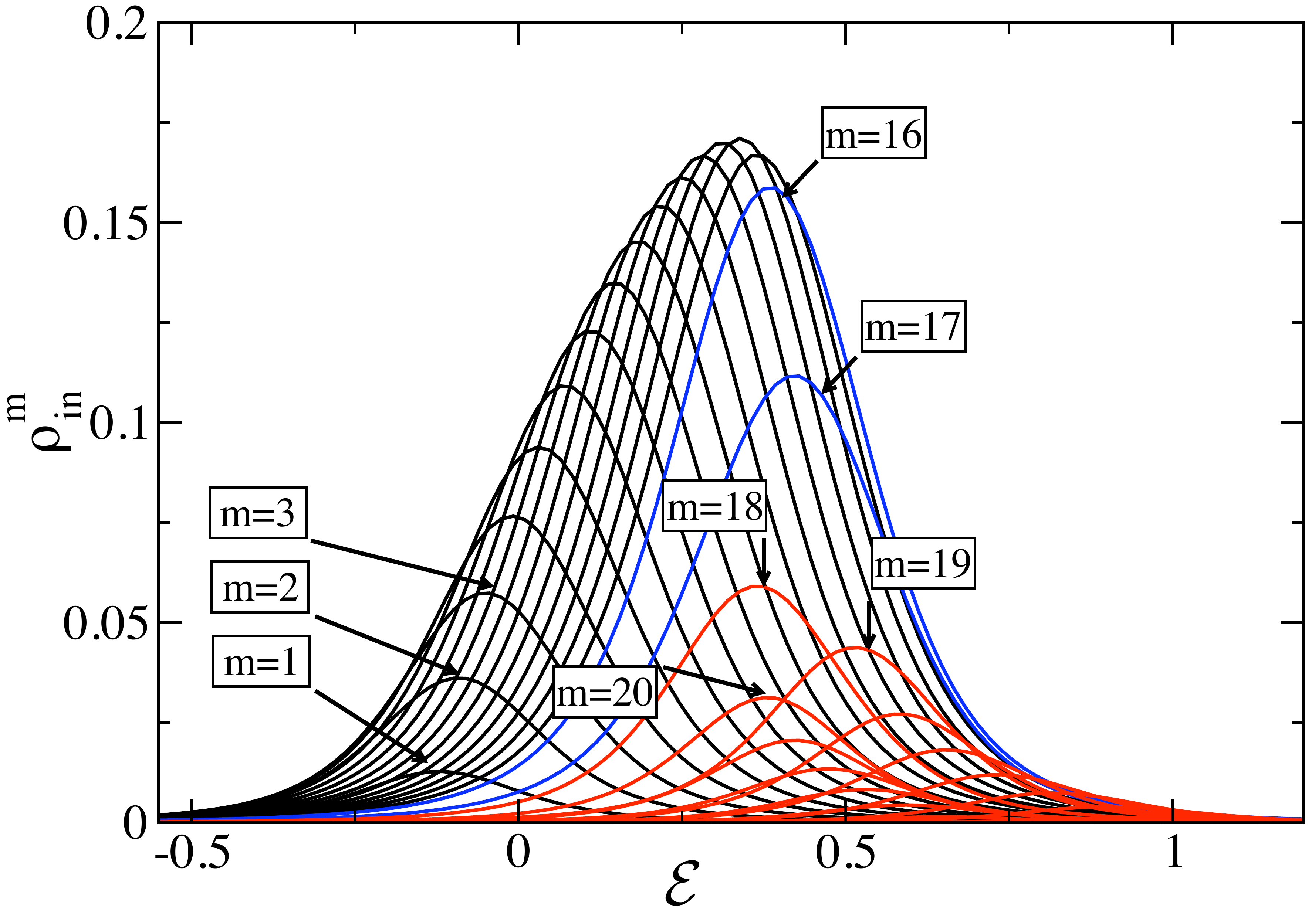}}
\caption{(color online). 
Inelastic cross section $\rho_{\rm in}^n$  versus the total energy ${\cal E}$, in units of the tunneling strength $k$. For better visibility we chose $N=30$ particles but
otherwise same parameters as in Fig.~\ref{fig.PN}.
Some representative channels are labeled by their index $m$. The black (red) curves belong to preparations of the BEC
in the Rabi (self-trapped) region of the spectrum, while the blue curves correspond to initial preparations in
eigenstates with energies close to the separatrix energy.
}
\label{fig.rho_in}
\end{figure}

In Fig.~\ref{fig.rho_in}, we plot the inelastic cross section $\rho_{\rm in}^m({\cal E})$ as a function of the total energy ${\cal E}$, for all $m$.
That is, each curve corresponds to an initial preparation of the BEC in a different energy eigenstate $|E_m\rangle$. 
We use the same parameters as in Fig.~\ref{fig.PN} and -- for the sake of clarity -- choose a smaller system with $N=30$ bosons. 
As the channel number increases from $m=1$ to $m=15$, the maximum of $\rho_{\rm in}^m({\cal E})$ -- referred to in the following as
the {\em resonance position} -- monotonically shifts to larger energy values (solid black lines), until, around channels $m=16$ and $m=17$
(solid blue lines), the behavior changes. 
For channels $m \ge 18 $ (red lines), the resonance position oscillates, i.e., for $\rho^{18}_{\rm in}$ it is located at ${\cal E}=0.36$,
$\rho^{19}_{\rm in}$ becomes maximal at ${\cal E}=0.51$, while $\rho^{20}_{\rm in}$ takes its maximal value at ${\cal E}=0.38$, and so on.

This ``splitting effect'' is the most drastic consequence of the underlying mean-field dynamics.
In Fig.~\ref{fig.rho_in}, black curves correspond to
the Rabi region, while red curves refer to self-trapping.
Quite intuitively, one relates the oscillations of the resonance positions to the alternating mean
occupation $Q_{nn}$ 
(see right panel in Fig.~\ref{fig.Q-matrix}).
In other words, for a {\em pair} of curves (like e.g. $\rho^{18}_{\rm in}$ and $\rho^{19}_{\rm in}$), the one with
lower (higher) resonance position corresponds to a self-trapped state localized in well one (two).

This intuition is corroborated by a simplified expression for $\rho^m_{\rm in}({\cal E})$, that we derive from
first-order perturbation theory with respect to the probe-target interaction strength $\alpha$ (see Appendix B),
and which holds for the intermediate values of probe-target interaction $\alpha$ and coupling parameter $\gamma$ considered here:
\begin{eqnarray}
\rho^m_{\rm in,Born}({\cal E}) &\simeq &2 {v}_m^{-1} \frac{\alpha^2 { v}_m^2}{[(1-\gamma) ({\cal E} - E_m) - \alpha Q_{mm} ]^2 + \gamma^2 { v}_m^2} \nonumber \\ 
&& \times  \left[ \langle E_m|\hat{n}^2_1|E_m\rangle  - \langle E_m| \hat{n}_1|E_m\rangle^2 \right].
\label{eq.rho_in_formula}
\end{eqnarray}
Neglecting for a moment the (weak) energy dependence of the velocity $v_m$, Eq.~(\ref{eq.rho_in_formula})
takes a Lorentzian shape.
Accordingly, the resonance positions are well approximated by $E_n+\alpha/(1-\gamma) Q_{nn}$.
In the Rabi region, they thus experience a constant shift due to the approximately constant spacing of the BH energies $E_n$
(see right panel of Fig.~\ref{fig.separatrix}).
In contrast, self-trapped states are nearly pairwise degenerate and the spacing between adjacent pairs grows linearly.
In this regime, the oscillation in the occupation number $Q_{nn}$ causes the splitting of the resonance positions.

We note that the only ingredients in (\ref{eq.rho_in_formula}) are the eigen\-energy $E_m$ of the BH system, as well as the
expectation value $Q_{mm}$ and variance $\sigma_m^2$ of the number counting operator $\hat n_1$ in the corresponding
eigenstate $|E_m\rangle$.
In comparison with the standard deviation $\sigma_n$ shown in Fig.~\ref{fig.Q-matrix}, expression (\ref{eq.rho_in_formula}) reproduces
quite well the overall trend of the inelastic cross section $\rho^m_{\rm in}({\cal E})$:
In the Rabi as well as in the self-trapped region,  $\rho^m_{\rm in}({\cal E})$ assumes its largest (smallest) values for initial
target energies close to (far from) the separatrix energy $E_{\rm sep}$.

%%%%%%%%%%%%%%%%%%%%%%%%%%%%%%%%%%%%%%%%%%%%%%%%%%%%%%%%%%%%%%%%
%%%%%%%%%%%%%%%%%%%%%%%%%%%%%%%%%%%%%%%%%%%%%%%%%%%%%%%%%%%%%%%%

\section{Conclusion}
The properties of a BEC trapped in a double-well potential were analyzed via the inelastic,
quantum mechanical scattering of a probe particle. 
Traces of the underlying mean-field phase space, like the appearance of self-trapped
solutions, were unambiguously identified in various, experimentally accessible quantities.
Based only on the expectation value and variance of the probe-target interaction operator,
an analytical expression was derived that elucidates the main observations.
Finally, the proposed scattering setup represents a {\em non-destructive} measurement of the condensate, what is
in contrast to standard techniques, like time-of-flight imaging, that necessarily result in the destruction of the BEC.

\begin{acknowledgement}
We acknowledge financial support by DFG {\em Research Unit 760} and
the US-Israel Binational Science Foundation (BSF), Jerusa\-lem, Israel, and by a grant from
AFOSR No. FA 9550-10-1-0433.
\end{acknowledgement}

%%%%%%%%%%%%%%%%%%%%%%%%%%%%%%%%%%%%%%%%%%%%%%%%%%%%%%%%%%%%%%%%
%%%%%%%%%%%%%%%%%%%%%%%%%%%%%%%%%%%%%%%%%%%%%%%%%%%%%%%%%%%%%%%%

\appendix

\section{Parity in the Rabi region of the spectrum}

In the Rabi region and for vanishing potential bias, the energy eigenstates $|E_m\rangle$ (with $E_m < E_{\rm sep}$) of the BH Hamiltonian have well-defined parity,
i.e. ${\hat P}|E_m\rangle = \pm |E_m\rangle$, with ${\hat P}$ the parity operator that interchanges the particles between
the two wells.
We now show that the interaction operator $\hat n_1$ cannot induce transitions between energy eigenstates with same parity.

We first calculate the commutator $[{\hat P},\hat n_1]$. 
in the Fock basis $|n_1,N-n_1\rangle \equiv |n_1\rangle$, where $n_1$ denotes the number of
bosons in well one. In this basis, $\hat{P}|n_1\rangle = |N-n_1\rangle$, and we have
\begin{eqnarray}
[{\hat P}, {\hat n}_1] |n_1\rangle
 & = &  ( n_1 {\hat P} - (N-n_1) {\hat P} ) |n_1\rangle \nonumber \\
 & = &    (2{\hat n}_2 - N )  {\hat P}  |n_1\rangle.
\end{eqnarray}
Assuming that the two energy eigenstates $|E_m\rangle$ and $|E_n\rangle$ ($m\neq n$) have {\em same} parity, we calculate the
corresponding off-diagonal element of ${\hat n}_1$:
\begin{eqnarray}
\langle E_n | {\hat n}_1 | E_m\rangle & = & \langle E_n | {\hat P} {\hat n}_1 {\hat P} | E_m\rangle  \nonumber \\
& = & \langle E_n |  (2{\hat n}_2 - N )  {\hat P} {\hat P}  + {\hat n}_1  {\hat P} {\hat P} | E_m\rangle \nonumber \\
& = & \langle E_n |  {\hat n}_2 | E_m\rangle \nonumber \\
& = & \langle E_n | N - \hat{n}_1 | E_m\rangle = - \langle E_n |  {\hat n}_1 | E_m\rangle  \, ,
\end{eqnarray}
i.e. $\langle E_n |  {\hat n}_1 | E_m\rangle = 0$.

%%%%%%%%%%%%%%%%%%%%%%%%%%%%%%%%%%%%%%%%%%%%%%%%%%%%%%%%%%%%%%%%

\section{Perturbation theory}

In this Appendix, we derive Eq.~(\ref{eq.rho_in_formula}). The starting point of this calculation is
the Born expansion of the $S_T$-matrix with respect to the perturbation parameter $\alpha$: 
We incorporate the diagonal of the $Q$-matrix,
\begin{equation}
\tilde{Q}_{nm} = \delta_{nm} Q_{nm} \, ,
\label{eq.cha_6_q_tilde}
\end{equation}
into the unperturbed Hamiltonian $H_{BH}$, and treat solely the non-diagonal part of the $Q$-matrix,
\begin{equation}
 \bar{Q} = Q - \tilde{Q},
\end{equation}
as the perturbation. 
Hence, we obtain as first-order Born approximation of (\ref{eq.S_T}) the following expression:
\begin{eqnarray}
\label{eq.S_T_Born}
S_{T,\rm{Born}} 
&=& i \, { v}^{1/2} 
\frac{\gamma}{ (1{-}\gamma) ({\cal E} -  H_{BH}) - \alpha \tilde{Q}   +  i\gamma { v} }
{ v}^{1/2}  \nonumber  \\
 &+& 
i\gamma \alpha\,
\frac{ { v}^{1/2}}{ (1{-}\gamma) ({\cal E} - H_{BH}) - \alpha \tilde{Q}  +   i\gamma { v}} \bar{Q}\nonumber \\
&\times&  \frac{ { v}^{1/2} }{ (1{-}\gamma) ({\cal E} - H_{BH}) - \alpha \tilde{Q} +  i\gamma { v} } \, .  \\
\nonumber
\end{eqnarray}
Note that the first term in (\ref{eq.S_T_Born}), termed $S_D$, is 
irrelevant for the inelastic cross section $\rho_{\rm in}$, since it is diagonal.
We can further use the perturbative result (\ref{eq.S_T_Born}), to obtain an analytical estimate for
the inelastic cross section and the transmission probability.
To this end, we rewrite
(\ref{eq.S_T_Born}) as:
\begin{eqnarray}
S_{T,\rm{Born}}     &=& S_D - i\kappa B
\label{eq.cha_6_S_D}
\end{eqnarray}
where $\kappa=\alpha/\gamma$ is the rescaled perturbation parameter and
$B \equiv S_D {v}^{-\frac{1}{2}} \bar{Q} {v}^{-\frac{1}{2}} S_D$ is the rescaled perturbation operator.
From Eqs.~(\ref{eq.rho_in}) and (\ref{eq.cha_6_S_D}) one obtains the inelastic cross section:
\begin{eqnarray}
\nonumber
\rho^m_{\rm in,\rm{Born}}
				&=& 2\kappa^2 \left[\sum_n  B_{mn}B_{nm}^* - B_{mm}B_{mm}^*\right]  \\[0.1cm]
			     	&=&  	2 \kappa^2 [\langle E_m| BB^{\dagger}| E_m\rangle - \langle E_m|B| E_m \rangle^2] \, . 
\label{eq.cha_6_varianz}
\end{eqnarray}
It is interesting to note that, in leading order of $\kappa$, $\rho^m_{\rm in, Born}$ depends only on 
the variance of the perturbation operator $B$ in the eigenstate $|E_m\rangle$ of the Bose-Hubbard Hamiltonian.\footnote{A similar
expression appears when one calculates the energy spreading of driven systems \cite{CK01,HKG09}.}
With the definition of $B$ and some basic algebra we obtain
\begin{eqnarray}                                                                                                                                                   
&& \rho^m_{\rm in,Born} = \nonumber\\
&&  2\kappa^2 { v}_m^{-1} \frac{\gamma^2 { v}_m^2}{[(1-\gamma) ({\cal E} - E_m) - \alpha Q_{mm} ]^2 + \gamma^2 { v}_m^2}  \\[0.1cm]
&&                                  \times \sum_k \bar{Q}_{mk} \underbrace{{ v}_k^{-1} \frac{\gamma^2 { v}_k^2}{[(1-\gamma) ({\cal E}  - E_k) - \alpha Q_{kk} ]^2 + \gamma^2 						{ v}_k^2}  }_{A}  \bar{Q}_{km} \nonumber \, .
\label{eq.cha_6_huhu}
\end{eqnarray}
In order to simplify the above formula, we assess the part $A$ in the last 
equation: We have numerically verified that for moderate values of $\alpha$ and $\gamma \le 0.2$,
each of the $A$-terms is equal or smaller than unity (apart from the ${v}_k^{-1}$ its a Lorentzian curve, i.e. we have to assure that 
${v}_k^{-1}\le 1$).
The inelastic cross section is then given by:
\begin{eqnarray}
\rho^m_{\rm in, Born} &\ge& 2 { v}_m^{-1} \frac{\alpha^2 { v}_m^2}{[(1-\gamma) ({\cal E} - E_m) - \alpha Q_{mm} ]^2 + \gamma^2 { v}_m^2} \nonumber \\
                                  && \times     \left[ \langle E_m|\hat{n}^2_1|E_m\rangle  - \langle E_m| \hat{n}_1|E_m\rangle^2 \right]  \, .
\label{eq.cha_6_rho_in_formula}
\end{eqnarray}

%%%%%%%%%%%%%%%%%%%%%%%%%%%%%%%%%%%%%%%%%%%%%%%%%%%%%%%%%%%%%%%%
%%%%%%%%%%%%%%%%%%%%%%%%%%%%%%%%%%%%%%%%%%%%%%%%%%%%%%%%%%%%%%%%

% BibTeX users please use
\bibliographystyle{epjc}
\bibliography{SOD}

\end{document}